\documentclass{aa}
\usepackage{graphics}

\begin{document}

   \thesaurus{01     
              (02.02.1;  
               13.07.3;  
               13.07.1;  
               13.07.2)}  
   \title{The Dyadosphere of Black Holes and Gamma-Ray Bursts}


   \author{Remo Ruffini}


   \institute{I.C.R.A.-International Center for Relativistic Astrophysics
and
Physics Department, University of Rome ``La Sapienza", I-00185 Rome,
Italy\\
            email: ruffini@icra.it
             }


   \maketitle

   \begin{abstract}

Works on the Dyadosphere are reviewed.

   \end{abstract}

%

I am proposing and give reasons that with Gamma Ray Bursts, for the first time we are witnessing,  in real time, the moment of gravitational collapse to a Black Hole. Even more important, the tremendous energies involved by the energetics of these sources, especially after the discoveries of their afterglows and their cosmological distances (Kulkarni {\it et. al.} 1998), clearly point to the necessity and give the  opportunity to use as an energy source of these objects the extractable energy of Black Holes.

That Black Holes can only be characterized by their mass-energy $E$, charge $Q$ and angular momentum $L$ has been advanced in a classical article (Ruffini \& Wheeler 1971), the proof of this has been  made  after twenty five years of meticulous mathematical work. 
One of the crucial points in the Physics of Black Holes  was to realize that energies comparable to their total mass-energy could be extracted from them. The computation of the first specific example of such an energy extraction process, by a gedanken experiment, was given in (Ruffini \& Wheeler 1970) and (Floyd and R. Penrose 1971) for the rotational energy extraction from a Kerr Black hole, see Figure (1).
The reason of showing this figure is not only to recall the first such explicit computation, but to emphasize how contrived and difficult such a mechanism can be: it can only work for very special parameters and should be in general associated to a reduction of the rest mass of the particle involved in the process. To slow down the rotation of a Black Hole and to increase its horizon by the accretion of counterrotating particles is almost trivial, but to extract the rotational energy from a Black Hole, namely to slow down the Black Hole {\it and}  keep its surface area constant, is extremely difficult, as clearly pointed out also by the example in Figure (1). The above  gedanken experiments, extended as well to electromagnetic interactions,  became of paramount importance not for their direct astrophysical significance but because they gave the tool for testing the physics of Black Holes and identifing their general mass-energy formula  (Christodoulou \& Ruffini 1971). The crucial point is that a transformation at constant surface area of the Black Hole, or reversible in the sense of ref.~\cite{rc},  could release an energy up to 29\% of the mass-energy of an extremal rotating Black Hole and up to 50\% of the mass-energy of an extremely magnetized and charged Black Hole.
\begin{figure}
\centering
\resizebox{\hsize}{8cm}{\includegraphics{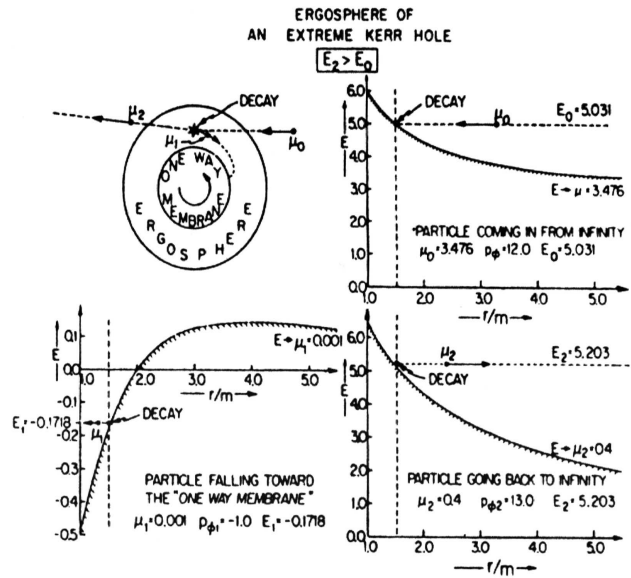}}
\caption{(Reproduced from Ruffini and Wheeler with their kind permission.) Decay of a particle of rest-plus-knetic energy $E_\circ$
into a particle which is captured into the black hole with positive
energy as judged locally, but negative energy $E_1$ as judged from      infinity, together with a particle of rest-plus-kinetic energy      $E_2>E_\circ$ which escapes to infinity. The cross-hatched curves give the effective potential (gravitational plus centrifugal) defined by the solution $E$ of Eq.(2) for constant values of $p_\phi$ and $\mu$ (figure and caption
reproduced from Christodoulou  1970.}
\label{pic2}
\end{figure}

%


Various models have been proposed in order to tap the rotational energy of Black Holes by the processes of relativistic magnetohydrodynamic. It is likely however that these processes are relevant over the very long time scales characteristic of the accretion processes.

In the present case of the Gamma Rays Bursts a prompt mechanism, on time scales shorter then a second, for depositing the entire energy in the fireball  at the moment of the triggering process of the burst, appears to be at work. For this reason we are here considering a more detailed  study of the vacuum polarization processes $à$  {\it a' la} Heisenberg-Euler-Schwinger (Heisenberg W. \& Euler H. 1931, Schwinger J. 1951)
around a Kerr-Newman Black Hole first introduced by Damour and Ruffini (Damour T. and Ruffini R., 1975).
The fundamental points  of this process can be simply summarized:

\begin{itemize}

\item They occur in an extended region arround the Black Hole, the Dyadosphere, extending from the horizon radius $r_+$ to the Dyadosphere radius $r_{ds}$ see (Preparata, Ruffini \& Xue 1998a,b). Only Black Holes with a mass larger than the upper limit of a neutron star and up to a maximum mass of $6\cdot 10^{5}M_\odot$ can have a Dyadosphere, see (Preparata, Ruffini \& Xue 1998a,b) for details.

\item The efficiency of transforming the mass-energy of Black Hole into particle-antiparticle pairs outside the horizon can approach 100\%, for Black Holes in the above mass range see  (Preparata, Ruffini \& Xue 1998a,b) for details. 

\item The pair created are mainly positron-electron pairs and their number is much larger than the quantity $Q/e$ one would have naively expected on the ground of qualitative considerations. It is actually given by $N_{\rm pairs}={Q\over e}(1+{r_{ds}\over \hbar/mc})$, where $m$ is the electron mass.  The energy of the pairs and consequently the  emission of the associated electromagnetic radiation peaks in the X-gamma rays region, as a function of the Black Hole mass.

\end{itemize}

I recall some of the results on the Dyadosphere.
We consider the collapse to almost general Black Hole endowed with an electromagnetic field (EMBH). Following Preparata, Ruffini \& Xue (1998a,b), for simplicity we consider the case of a non rotating Reissner-Nordstrom EMBH to illustrate the basic gravitational-electrodynamical process. 
The number density of pairs created in the Dyadosphere is 
\begin{equation}
N_{e^+e^-}\simeq {Q-Q_c\over e}\left[1+{
(r_{ds}-r_+)\over {\hbar\over mc}}\right],
\label{n}
\end{equation}
where
\begin{equation}
r_{\rm ds}=\left({\hbar\over mc}\right)^{1\over2}
\left({GM\over c^2}\right)^{1\over2} 
\left({m_{\rm p}\over m}\right)^{1\over2}
\left({e\over q_{\rm p}}\right)^{1\over2}
\left({Q\over\sqrt{G} M}\right)^{1\over2}.
\label{rds}
\end{equation} 
Due to the very large pair density 
and to the sizes of the
cross-sections for the process $e^+e^-\leftrightarrow \gamma+\gamma$, 
the system is expected to thermalize to a plasma configuration for which
\begin{equation}
N_{e^+}=N_{e^-}=N_{\gamma}=N_{\rm pair}
\label{plasma}
\end{equation}
and reach an average temperature
\begin{equation}
kT_\circ={ E^{\rm tot}_{e^+e^-}\over3N_{\rm pair}\cdot2.7},
\label{t}
\end{equation}
where $k$ is Boltzmann's constant.
The discussions on the relativistic expansion of the Dyadosphere are presented in a separated paper (see e.g.~Ruffini, Salmonson, Wilson \& Xue 1999 in this proceeding).   

Before concluding I would like to return to the suggestion, advanced by Damour and Ruffini, that a discharged EMBH can be still extremely interesting from an energetic point of view and responsible for the acceleration of ultrahigh energy cosmic rays. I would like just to formalize this point with a few equations: It is clear that no matter what the initial conditions leading to the formation of the EMBH are, the final outcome after the tremendous expulsion of the PEM pulse will be precisely a  Kerr Newman solution with a critical value of the charge. If  the Background metric has a Killing Vector, the scalar product of the Killing vector and the generalized momentum 
\begin{equation}
P_\alpha = m U_\alpha + e A_\alpha,
\label{gm}
\end{equation}
is a constant along the trajectory of any charged gravitating particle following the relativistic equation of motion in the background metric and electromagnetic field (Jantzen and Ruffini 1999). Consequently an electron (positron) starting at rest in the Dyadosphere will reach infinity with an energy $E_{kinetic}\sim 2mc^2({GM\over c^2})/({\hbar\over mc})\sim 10^{22}$eV for $M=10M_\odot$.

\end{document}